\documentclass[12pt, british]{article}
\usepackage[a4paper]{geometry}
\usepackage{graphicx, listings, setspace, hyperref, verbatim}
\usepackage{amsmath, amssymb, mathtools, esint}
\usepackage{tikz, tabu, tabularx, makecell, gensymb, adjustbox, subcaption, float}
\usetikzlibrary{shapes, shadows, arrows}
\usepackage{textcomp}
\usepackage{mathptmx}
\usepackage{microtype}
\usepackage[T1]{fontenc}
\usepackage[utf8]{inputenc}
\usepackage{mathptmx}
\usepackage{braket}
\usepackage{hyperref}
\hypersetup{colorlinks=true,linkcolor=blue,citecolor=cyan}
\usepackage{authblk}

\usepackage[british]{babel}
\usepackage[babel=true]{csquotes}
\usepackage[backend=biber, style=ext-authoryear,  maxcitenames=2, dashed=false, url=false, doi=false, eprint=true, giveninits=true, uniquename=init, innamebeforetitle=true, maxbibnames=6]{biblatex} 

\AtEveryBibitem{%
 \clearfield{note}%
  \clearfield{series}%
 \clearfield{number}%
 \clearfield{pagetotal}%
 \clearfield{chapter}%
  }

\begin{document}

\title{Unpacking Black Hole Complementarity}
\author{Siddharth Muthukrishnan}
\date{June 2023}
\affil{Department of History and Philosophy of Science, University of Pittsburgh\\  \href{mailto:siddharth@pitt.edu}{siddharth@pitt.edu} \\
{\small Forthcoming in \textit{The British Journal for the Philosophy
of Science}}}

\maketitle 




\begin{abstract}
\noindent To what extent does the black hole information paradox lead to violations of quantum mechanics? I explain how black hole complementarity provides a framework to articulate how quantum characterizations of black holes can remain consistent despite the information paradox. I point out that there are two ways to cash out the notion of consistency in play here: an operational notion and a descriptive notion. These two ways of thinking about consistency lead to (at least) two principles of black hole complementarity: an operational principle and a descriptive principle. Our background philosophy of science regarding realism/instrumentalism might initially lead us to prefer one principle over the other. However, the recent physics literature, which applies tools from quantum information theory and quantum computational complexity theory to various thought experiments involving quantum systems in or around black holes, implies that the operational principle is successful where the descriptive principle is not. This then lets us see that for operationalists the black hole information paradox might no longer be pressing.
\end{abstract}



\onehalfspacing

\section{Introduction}

The black hole information paradox identifies a pivotal and productive tension within quantum mechanical treatments of black holes. Something seems to go wrong if a black hole is a quantum statistical object with its near-horizon physics describable with quantum field theory. Because the paradox combines quantum mechanics and gravity while remaining in a regime with good theoretical control, it has become a central problem in modern theoretical physics.\footnote{See, for example, \parencites{Belot1999-BELTHI-2, wallace_2020, marolf2017black} for systematic overviews.} 

`Black hole complementarity' is a label for an influential set of ideas~\parencites{PhysRevD.48.3743, PhysRevD.52.6997, almheiri2013black,Hayden_2007, RevModPhys.88.015002} that respond to the paradox. Many claims come under that label in the literature, so it can be confusing what an appeal to `black hole complementarity' is meant to do. I take the governing thought behind these appeals to be that black hole complementarity is a way to sand off the bite of the information paradox, allowing one to argue that the paradox does not ramify up into a physically or an empirically relevant inconsistency. This has precedent in the history of science, cases where an inconsistency is not a death knell for a theory but merely an obstacle to work around, a part of the theory one handles with care.\footnote{See, for example,~\parencites{vickers2013understanding} for several examples of inconsistent but successful scientific theories.} 

I will articulate and distinguish two principles of black hole complementarity embedded in these discussions: an operational principle and a descriptive principle. The operational principle says that any experiment conducted by an observer in or near an evaporating black hole will always be consistent with quantum mechanics. The principle is operational because it makes essential appeal to what is empirically accessible to observers.\footnote{I use the term `operational' as it is used in the physics literature on black hole complementarity. This isn't the way Percy Bridgman~\parencites{sep-operationalism} used it. Nor is it the sense common in information theory, wherein to operationalize a quantity (for example, Shannon entropy) is to specify a task for which the quantity quantifies the efficiency of the task (such as message compression).} The descriptive principle says that the infalling and exterior descriptions of the physics of an evaporating black hole are consistent descriptions of the same physics.  

Our background philosophy of science provides \emph{prima facie} support for one or the other principle. Realists, who think theories should aim to describe the world, will likely lean towards the descriptive principle since it constrains realistically interpretable descriptions. Instrumentalists, who are fine with theories being merely useful instruments for explanation and prediction, will likely have no qualms with the operational principle, since an instrument meant to help observers might refer to observers. However, these inclinations are defeasible by evidence. I will argue the recent physics literature indicates the operational principle succeeds and the descriptive principle fails. More precisely, the physics of quantum black holes seems to describe scenarios that violate quantum mechanics, leading to the failure of descriptive complementarity.  However, these violations can't be operationalized---no single observer can see a violation of quantum mechanics. Consequently, one must at least tentatively adopt operational complementarity. 

To be clear: operational and descriptive complementarity aren't necessarily opposed; the literature isn't divided into two camps endorsing one or the other attitude towards black hole complementarity. Often those who talk in descriptive terms in some cases will switch to operational talk in other cases. After all, descriptive complementarity entails operational complementarity: if our theories describe the world as consistent with quantum mechanics, then no observer will see a violation of quantum mechanics. The tension I highlight, rather, is that operational complementarity has much support in scenarios where descriptive complementarity seems to fail. 

Where does one go from there? For the realistically inclined, the operational principle is to be explained by future descriptive physics. But if one is comfortable with operationalism, then the information paradox plausibly dissolves away. 

Here's the plan. First, I briefly review the black hole information paradox, focusing on the Page-time paradox (section~\ref{sec:bhip}).  Next, I introduce black hole complementarity, and articulate operational and descriptive complementarity (section~\ref{sec:bhc}). I then discuss three families of thought experiments, each attempting to identify contradictions in black hole physics. The first is about horizon crossers (section~\ref{sec:horizoncross}). Here, I argue operational and descriptive complementarity are both successful. The second concerns a potential violation of quantum no-cloning (section~\ref{sec:clone}). Here, descriptive complementarity fails while operational complementarity succeeds. Finally, I study a potential violation of entanglement monogamy (section~\ref{sec:monogamy}). Here too, descriptive complementarity fails but operational complementarity succeeds. I conclude with lessons for the information paradox (section~\ref{sec:concl}).

To be broadly accessible, I avoid technical details.  For a quick review of requisite background, see the Supplemental Material in \parencites{RevModPhys.88.015002}.

\section{The black hole information paradox}\label{sec:bhip}
                                                                
\begin{figure}[!t]
\centering
\includegraphics[width=0.3\columnwidth]{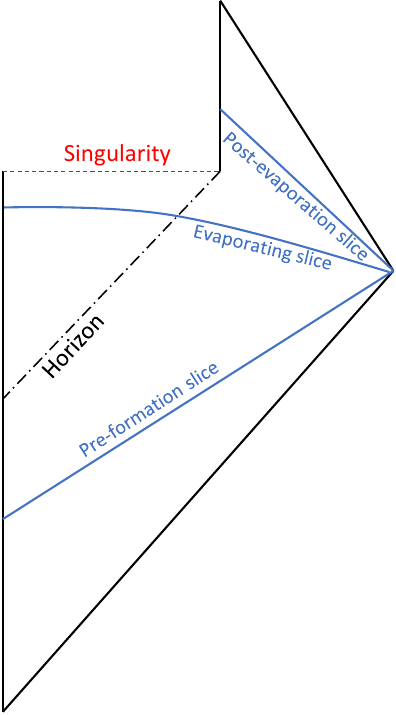}
\caption{A Penrose diagram for a black hole that forms and then evaporates away. (See, for example,~\textcites[pp. 4-14]{strominger1995houches} for how to draw and interpret Penrose diagrams.)}
\label{fig:evap_penrose}
\end{figure}

Consider a black hole that forms from matter collapse and then evaporates via Hawking radiation \parencites{hawking1975particle}. (See Figure~\ref{fig:evap_penrose}.) At least two paradoxes can be identified here: the total evaporation paradox and the Page-time paradox~\parencites{wallace_2020}. (```The'' black hole information paradox' is somewhat of a misnomer.)  This paper focuses on the Page-time paradox because black hole complementarity only makes sense as a response to the Page-time paradox. Moreover, the total evaporation `paradox' isn't nearly as compelling as the Page-time paradox \parencites{PhysRevD.14.2460, Mathur_2009, Unruh_2017, maudlin2017information,wallace_2020}.

Let me take a moment to clarify the term `information'. A dynamics is information-preserving if and only if past states of physical systems can be uniquely reconstructed from future states. Unitary dynamics is information-preserving. So talk about `information falling in' or `coming out' means certain physical systems are falling in or coming out---systems whose degrees of freedom are in states which, when taken along with other relevant physical systems, allow reconstruction of the past. `Information being lost' means physical systems have evolved into states that no longer allow for  reconstruction of earlier states. And `recovering information' could either mean obtaining physical systems whose states would allow for past reconstruction or mean the actual process of reconstructing earlier states using such systems---context will disambiguate.

\subsection{Page-time paradox}\label{subsec:page}

The Page-time paradox \parencites{PhysRevLett.71.3743, wallace_2020} presents a compelling argument to a seemingly absurd conclusion, and hence is a paradox in~\textcites{quine1962paradox}'s sense. The argument is compelling because its premises rely on physics in regimes believed to be under good theoretical control. Consequently, the Page-time paradox has dominated recent discussion by  physicists \parencites{Mathur_2009, polchinski2015bhip, RevModPhys.88.015002, marolf2017black}. 

According to the Page-time paradox, we get a contradiction between three statements: (A) the evaporation process is unitary;  (B) the black hole is a quantum statistical mechanical system with its von Neumann entropy (that is, its fine-grained entropy) bounded above by its Bekenstein-Hawking entropy (that is, its microcanonical entropy); and (C) Hawking radiation is perfectly thermal throughout evaporation, as predicted by \textcites{hawking1975particle}.

I won't argue here for the plausibility of these statements; for an extended defence see~\parencite{Wallace2018-WALTCF-9,Wallace2018-WALTCF-10,wallace_2020}. But a comment about the unitarity assumption (A): some may question its plausibility since certain proposed solutions to the measurement problem violate unitarity~\parencites{sep-qm-collapse}. If the world contains real physical collapses, then worrying about the unitarity of black holes is moot. But unitarity-preserving solutions (Everettian, Bohmian, neo-Copenhagenian) to the measurement problem are still viable, and so the information paradox is well worth discussing. Henceforth, I bracket the measurement problem. We will work within a unitary paradigm,  and we discuss unitarity of black hole physics in the same way one discusses unitarity of terrestrial quantum systems, such as quantum computers: We will ask whether the system's behaviour prior to measurement can be accounted for unitarily. This way, we can attribute unitarity to a system despite observers seeing definite outcomes for measurements (which is the case on all solutions to the measurement problem).

Let's bring out the paradox. (A) entails that the von Neumann entropy of radiation from a black hole equals the von Neumann entropy of the black hole. Why? At any given time, partition the total quantum system into two subsystems: radiation and black hole. We can assume the total state is pure since unitary evolution preserves purity. For pure states, the von Neumann entropy of the partitions must be equal---this quantity is the entanglement entropy: the larger this entropy, the larger the entanglement between the partitions. This along with (C) entails that black hole entropy increases throughout evaporation because perfectly thermal quanta are radiated. However, the Bekenstein-Hawking entropy of the black hole, which is proportional to the area of the horizon, will decrease as evaporation proceeds. Consequently, (B) entails that the black hole's von Neumann entropy starts decreasing at some point in time so as to remain smaller than the Bekenstein-Hawking entropy. Thus, (A), (B), and (C) together imply that the black hole's von Neumann entropy keeps increasing throughout evaporation and starts decreasing at some point during evaporation. Contradiction!

\begin{figure}[!t]
\centering
\includegraphics[width=0.5\columnwidth]{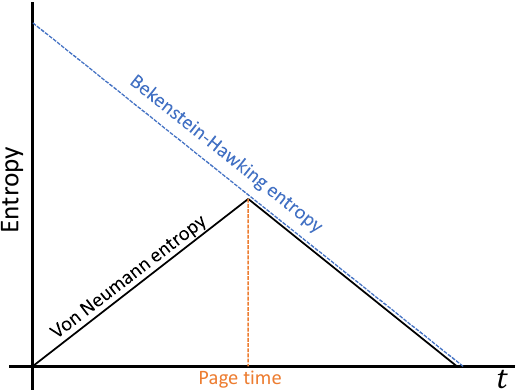}
\caption{The Page curve for the entropy of an evaporating black hole. The Page time is the time at which the von Neumann entropy of the black hole (or the radiation) has to start decreasing so as to not exceed the Bekenstein-Hawking bound.}
\label{fig:pagecurve}
\end{figure}

More quantitatively,~\textcites{PhysRevLett.71.3743} argued that the radiation from a unitarily evaporating quantum statistical black hole  will deviate from thermality starting roughly halfway through evaporation. This latter time is called the `Page time'. (See Figure~\ref{fig:pagecurve}.) Meanwhile, Hawking's argument tells us that the radiation will be thermal throughout evaporation, even past the Page time. Thus, the paradox.

The following stylized situation might help better understand the Page curve. Suppose the black hole forms from a large collection of highly entangled pairs of particles. Later, it radiates those particles one at a time, chosen at random. Each emitted particle will be in a highly mixed state. Initially, for any particle that's being emitted, it's unlikely that the particle's entangled partner has already been radiated. But halfway through the emission process, it becomes more likely than not that the partner is already in the radiation, and so we can expect the total state of the radiation to start purifying. This tracks the Page curve: initially the radiation is mostly highly mixed states, and so high entropy, but after halfway through, these mixed states start finding their partners, leading to the radiation becoming purer, and the entropy decreasing. As the radiation becomes purer, the past becomes more retrodictable---more information is coming out.

The Page-time paradox is a compelling argument to an absurd conclusion---indeed, a contradiction. The argument is compelling because no reference was made to the singularity or to post-evaporation quantum states. Around, and well past the Page-time, time evolution between slices is perfectly well-defined and hence there is no problem with defining unitarity, regardless of whether the spacetime is globally hyperbolic. Also, the argument was made in a regime where we have no reason to expect Hawking's argument for the thermality of black hole radiation to fail. We only need quantum field theory to hold near the horizon. True it's quantum field theory in a globally curved spacetime, but the curvature at the horizon can be made small as you like. More broadly, all the elements of the Page-time paradox---Hawking radiation (in both photons and gravitons), black hole evaporation (that is, including back-reaction),  and a quantum statistical description of black holes can be situated within the framework of Low-Energy Quantum Gravity, which includes quantum field theory on curved spacetime and semiclassical gravity as limits~\parencites{WALLACE202231}. Consequently, the Page-time paradox has real bite.

\subsection{The stretched horizon}\label{sec:stretch}

The stretched horizon is a striking consequence of a unitary quantum statistical description (that is, assumptions (A) and (B)) of a black hole. It will feature in the discussion to follow, so let me introduce it here. A quantum statistical description of the black hole (that is, (B)) requires an effective field theory for the exterior which has an entropy at the horizon bounded above by the Bekenstein-Hawking entropy. Moreover, for a unitary description of the black hole (that is, (A)), the exterior field theory requires a boundary that is strictly above the horizon (since things that cross the horizon cannot re-emerge) which can absorb and re-emit information that it encounters. A boundary surface located one Planck length above the horizon satisfies these requirements. (This location is set by the demand that the entropy of the exterior field theory not exceed the Bekenstein-Hawking entropy at the horizon.) This is the stretched horizon. The stretched horizon is timelike unlike the true horizon, which is null.  Encoding infalling information onto the stretched horizon necessarily requires Planck-scale physics;\footnote{See, for example,~\parencite[pp. 9-11]{BANKS199521}.} nevertheless, the stretched horizon can maintain unitarity for low-energy exterior physics by the way it interacts with exterior physics. The stretched horizon is a real entity in the reference frame of observers hovering outside the black hole.\footnote{See \parencite[chapter 7]{susskind2005introduction} for a pedagogical introduction to the stretched horizon. \textcites{PhysRevD.48.3743} provided the first systematic treatment of the idea of a stretched horizon, but \parencite{THOOFT1985727} is an important precursor (see \parencites[pp. 18-9]{RevModPhys.88.015002} for a summary of 't Hooft's argument).} In the classical limit, it will have its own distinctive viscosity and electrical resistance; it will respond in a local way to external perturbations---it will radiate, carry electrical currents, and oscillate. 

In the classical limit, the stretched horizon becomes the membrane in the membrane paradigm of~\textcite{thorne1986black}.\footnote{\parencites{priceSciAm1988} is an accessible overview.} The membrane paradigm helped conceptualize black holes as astrophysical objects that can interact with their surrounding electrical and magnetic fields, much like stars. One can view the quantum stretched horizon as a quantization of the classical membrane.

With the stage set, let's turn to our main topic: black hole complementarity.

\section{Black hole complementarity}\label{sec:bhc}

The Page-time paradox points to a contradiction between a statistical mechanical application of quantum mechanics and a field-theoretic application of quantum mechanics, with both applications in regimes where we believe quantum mechanics works well. It is to assuage this contradiction that we now turn to black hole complementarity. 

You might think: A contradiction is not the sort of thing that one `assuages'; if your theory yields a contradiction, then so much the worse for your theory. But this isn't entirely satisfying. The history of science provides many examples of seemingly inconsistent theories that were nevertheless successful. Apart from the examples in \textcites{vickers2013understanding}, a useful example is quantum mechanics itself. At least on some ways of conceiving of the quantum measurement problem (for example,~\textcites{albert1994quantum}), we have a straightforward inconsistency within quantum mechanics. However, to the chagrin of many philosophers of physics, quantum mechanics has enjoyed great success despite this inconsistency. So, physics might tolerate an inconsistency, but only insofar as it cannot be subjected to direct empirical test. Thus, we should ask: Can we point to some way in which the Page-time paradox leads to a clear observable violation of the predictions of quantum mechanics? If not, then that is the sense in which we would have `assuaged' the contradiction.

How might we extract an observable violation of quantum mechanics from the Page-time paradox? A straightforward experiment suggests itself. Take Bob, an observer hovering outside a black hole,\footnote{Bob will frequently be contrasted with Alice, an intrepid observer free-falling into the black hole.} patiently collecting all the radiation coming from it.\footnote{It might help to imagine Bob as a spectacular experimentalist, who has built a Dyson sphere around the black hole which collects every radiated quanta, including gravitons.} If Bob sees the total state of the radiation deviate from thermality past the Page-time, consistent with the Page curve (Figure~\ref{fig:pagecurve}), then his observations would be consistent with unitary evaporation---and so he would conclude Hawking's calculation is incorrect and quantum field theory is inapplicable near the horizon. And if the total state doesn't deviate from thermality, as predicted by Hawking, then he would know that black holes don't evaporate unitarily, and that the Page curve is mistaken. Either way, we would have an empirical test of the contradiction.

But can this seemingly straightforward experiment be carried out by Bob? It seems not. The core difficulty is that entropy isn't an observable in quantum mechanics: there is no Hermitian operator corresponding to entropy. Eigenstates of Hermitian operators are pure states, which are zero entropy; only mixed states have non-zero entropy. So Bob can't just measure the entropy and see whether he's in the Page scenario or the Hawking scenario.

Perhaps there's some clever way to experimentally distinguish the two scenarios? Consider the following. Say Bob knows the state on the pre-formation slice was the pure state $\ket{\psi}$. Further suppose he knows that if black hole evaporation is unitary, then it must be a particular unitary $U_\mathrm{bh}$. Then, once the black hole finishes (or almost finishes) evaporating, he can perform the two-outcome measurement $\{U_\mathrm{bh}\ket{\psi}\bra{\psi}U_\mathrm{bh}^\dagger, I - U_\mathrm{bh}\ket{\psi}\bra{\psi}U_\mathrm{bh}^\dagger\}$ on the radiation. If black hole evaporation were unitary, and so the Page curve were true, then he should get the first outcome with near certainty. If, however, the Hawking scenario were true, then very likely he'd get the second outcome.

Victory? No. The method above has a serious shortcoming: it requires that, if evaporation were unitary, then Bob can know what the black hole unitary $U_\mathrm{bh}$ is. And this is an implausible assumption. For one, it would be a rather strange epistemic situation to be in: to simultaneously be uncertain whether a black hole evaporates unitarily while being certain that if it did, then this is the precise unitary it would evolve under. For another, an adequate model of $U_\mathrm{bh}$ would require modeling Planck-scale physics, because from Bob's perspective, the dynamical processes governing emission of black hole radiation involve stretched horizon degrees of freedom, and as noted in section~\ref{sec:stretch}, those degrees of freedom are Planck-scale.

But perhaps Bob can do a similar experiment without a precise model of the black hole unitary---perhaps a good approximate and effective model would do? After all, much statistical physics can be tested without precise knowledge of microphysics. Unfortunately, however, the most plausible effective models of black holes are no better at helping resolve between the two entropy curves. Such models take a black hole to be well-modeled by a unitary transformation chosen uniformly at random ~\parencites{sekino2008fast,maldacena2016bound}. According to such models, the resulting state of the black hole and the emitted radiation after a significant period of evolution looks like a randomly chosen state in the total Hilbert space.\footnote{Given a Hilbert space, there is a unique uniform measure on the space of unitary operators acting on the Hilbert space---the so-called Haar measure (see, for example,~\parencites{fisher2023}). Using this measure we have a well-defined notion of a uniformly randomly chosen unitary transformation, and consequently a uniformly randomly chosen state in the Hilbert space.} Now, if we consider an arbitrary subsystem of a random pure state, then it is going to be nearly indistinguishable by measurements from a maximally mixed state (or: a thermal state at a certain temperature under an energy-conservation constraint) unless the subsystem is almost as big as the whole system~\parencites{popescu2006entanglement,hsu2009black,wang2017hawking}. It's intuitive to see why: If you are handed a randomly chosen vector in a Hilbert space, and you know nothing specific about the vector apart from how it was chosen, then, from your perspective, you have a maximally mixed state. And any non-special subsystem of a maximally mixed state will also look maximally mixed (or will look thermal if you have an energy constraint). Consequently, according to such models, the radiation subsystem will look indistinguishable from a highly mixed (high-entropy) state even if black hole evaporation is unitary, and the global state is pure. 

You might not convinced by the above argument; there must be some cleverer way to empirically display the contradiction identified by the Page-time paradox, no? In much of the rest of this paper, we will discuss attempts to show that the Page-time paradox results in a single-observer violation of quantum mechanics---and how those attempts fail. That such attempts will continue to fail is the import of what I call `operational black hole complementarity'. 

You might be suspicious of appeals to what observers like Bob can or cannot do. Below, I will explore whether such an appeal is dispensable, and how one might attempt dispensing with such appeals. What I call `descriptive complementarity' will arise from such an attempt.

\subsection{Operational and descriptive complementarity}\label{subsec:opdescpcomp}

I have introduced black hole complementarity as an operational principle, that is, a principle that makes essential, ineliminable reference to the capabilities of observers. After all, I was concerned with what sorts of experiments Bob, a particular observer, can or cannot do. Let's make explicit this way of thinking:
\begin{quote}
Operational complementarity: No experiment attempting to create the observation of a direct contradiction of the rules of quantum mechanics by a single observer near, or in, black holes will succeed.
\end{quote}

What do we mean by `observer' here? Obviously, we don't mean a human, though that might be suggested by  `Alice' and `Bob'. After all, humans can't plausibly survive and experiment in the extreme regimes near black holes. Rather, an `observer' here is some kind of experimental system that can collect and process quantum mechanical data. (Note that acquiring such data may also require being able to intervene on relevant quantum degrees of freedom to some extent. So really the notion at play here is that of an observer-agent, though I will simply use the term `observer'.) So these systems might be better imagined as some exquisitely sensitive automated instrument rather than a human. Nevertheless, we'll stick with the human-name convention common in this literature. 

There is one important constraint on what we consider as observers: We will assume that observers are only able to probe or control physics that is above the Planck scale. The main reason for this restriction is that we don't know how to coherently conceive of physical systems that can acquire fine-grained information about, and control over, Planck-scale degrees of freedom so as to count as observers.

Let me elaborate. Consider an analogy. The proscription against energy-conserving perpetual motion machines (that is, those that violate the second law of thermodynamics but not the first) can be seen as an operational restriction: observers can't construct and maintain those types of machines. This proscription is clearer and less controversial if we exclude observers of the Maxwell-demon-type. That is, if observers could observe and manipulate atoms and molecules with precision (without entropy gain) then that seems to make perpetual motion machines possible.  But whether this is a real possibility depends sensitively on the details of the constitution of atomic scale observers---for example, such observers might be buffeted around too much to sustainably carry out their interventions~\parencites{norton2013}. Thus, it is reasonable to exclude such observers in analyses that don't engage with detailed microphysics.\footnote{This way of thinking about the second law---that is, as a principle whose validity is contingent on the abilities of microphysically constituted observers---is close to Maxwell's own way of thinking about the principle~\parencites{MYRVOLD2011237}.} Similarly, in analysing operational complementarity, we exclude observers who can access Planck-scale physics. Imposing such a restriction simply means that the validity of operational complementarity is left open in Planck-scale regimes: it may or may not continue to hold at, and beyond, that regime. We leave it open because we cannot presently analyse the plausibility of the principle in that regime. Indeed, analysing operational complementarity at the Planck scale might require going well beyond the conceptual affordances of our current best theories---the affordances that come with concepts such as `spacetime', `particles', and `fields'. And in the absence of a Planck-scale theory, we don't know what alternative conceptual tools to employ (compare footnote~\ref{fn:sptbreak}).

A related, but different, justification for excluding Planck-scale observers flows from the fact that the Page-time paradox is pressing precisely because its premises are expected to hold in a regime where we think Planck-scale physics should be irrelevant. It is then natural to ask whether one can empirically probe the paradox without requiring access to resources that go well beyond this regime. For, if it were true, as operational complementarity suggests, that problematic consequences of the Page-time paradox would not manifest in the above-Planck-scale regime, then that would go some way towards assuaging paradox, because it would mean the relevant regime is under at least some kind of control, namely an empirical one. Thus, insofar as we are considering black hole complementarity as a way of responding to the information paradox, it is reasonable to restrict the response to the same regime as the one in which the paradox arises.\footnote{Compare~\textcites{Bokulich2005-BOKDBH} for related observations, though not specifically in the context of operational ways of thinking about black hole complementarity.}

Thus, in what follows, I will consider physics at the Planck scale as inaccessible to observers. I  leave the notion of `observer' underspecified beyond this point. Once we start considering examples, it should be clearer what sorts of things are observers.

Something like operational complementarity is prevalent in the recent physics literature, as evidenced by some quotes: 
\begin{itemize}
\item \textcites[p. 2]{Hayden_2007}: `...``black hole complementarity,'' according to
which no violations of the accepted principles of quantum physics can be detected by any observer,
whether outside or inside the black hole.'
\item \textcites[p. 1]{bousso2013}: `Complementarity distinguishes the viewpoint of an observer who
remains far from the black hole, Bob, from that of an infalling observer, Alice. These viewpoints have to be consistent as long as they can be operationally compared.'
\item \textcites[p. 1]{nomura2013complementarity}: `Black hole complementarity asserts that there is no contradiction between the two pictures [that is, the infalling and the exterior], since the statements by the two observers cannot be operationally compared...'
\end{itemize}

One may welcome such operational principles, especially if one is already sympathetic to some kind of instrumentalism about scientific theories. For, if one thinks that our scientific theories are essentially predictive tools, then it shouldn't be surprising that some aspects of our theories make essential reference to the capabilities of observers---after all, guiding observers is what theories are for. And such a view isn't at all alien in the context of quantum mechanics. If one prefers an instrumentalist response to the quantum measurement problem, then an instrumentalist principle in the context of the quantum physics of black holes is just so much more grist for one's mill.\footnote{A classic statement of an instrumentalist view on quantum mechanics is by \textcites[p. 70]{fuchs2000quantum}: `...quantum theory does not describe physical reality. What it does is provide an algorithm for computing probabilities for the macroscopic events (``detector clicks'') that are the consequences of our experimental interventions.'}

Typically however, in philosophy of physics, and in philosophy of science more broadly, there is a wariness surrounding operational principles since it's unclear how principles that refer to observers could be interpreted as describing a mind-independent reality.  This is in keeping with the popularity of scientific realism in these communities.\footnote{The 2020 Philpapers Survey found that about 60\%  of the survey's target faculty in philosophy of science accept or lean towards scientific realism, with that fraction rising to about 70\% of the target faculty in philosophy of physical science~\parencites{BourgetManuscript-BOUPOP-3}.}

The relative antipathy towards instrumentalist views in broader philosophy of science is related to the difficulties of making work a verificationist semantics for scientific theories, as was attempted by the logical empiricists.\footnote{A particularly influential critique of verificationism was advanced by~\textcites[section V]{Quine1951-QUITDO-3}.} These difficulties constitute an important reason why many philosophers of science are realists, but it is worth remarking that such difficulties are even persuasive to anti-realists such as \textcites{van1980scientific}, who makes it a point to avoid cashing out the content of scientific theories in terms of what is observable (indeed, he thinks scientific theories ought to be construed literally)---it's only that we need not take them to be true. Thus, even some anti-realists will be unfriendly to operational complementarity.

Within philosophy of physics specifically, the following argument has had significant purchase: Observers are physical systems, constituted by atoms and the like, and thus what they can or cannot do should ultimately be explicable in terms of observer-free physics. Hence, any principle that takes observers as primitive incurs the burden of explaining why we cannot state the capabilities of observers in observer-free terms given their physical constitution.\footnote{This is a version of an argument advanced by~\textcites{bell1990against} in the context of the quantum measurement problem.}

Given all this, philosophers may find the operationalism in operational complementarity unpalatable. And so one might ask whether the reference to observers in operational complementarity is really ineliminable. One might seek to eliminate the reference to observers and to state complementarity in a way that does not appeal to observers and their capabilities. I call such a way of stating complementarity `descriptive'. On this approach, instead of constraining what is possible for observers, we want to constrain our theories, interpreted as descriptions of the world. I will now supply one natural way of articulating a descriptive principle of complementarity. 

If we are to constrain descriptions, we need to first decide which descriptions we'll be constraining. Well, our goal is to eliminate the appeal to observers in our operational principle, so a natural way to generate the descriptions to be constrained would be the descriptions that attach to the two kinds of observers that often come up in this context---hovering and infalling. (This move is analogous to how in relativity, we can usually trade in talk of observers for talk of reference frames.) So the two different descriptions that attach to these two kinds of observers are:
\begin{itemize}
\item Exterior description---This is a consistent low-energy quantum mechanical description of the black hole and its exterior that includes degrees of freedom outside the horizon with the stretched horizon as the boundary. On this description, the black hole evolves unitarily and is a statistical-mechanical object with entropy bounded above by the Bekenstein-Hawking entropy. This description will be attributed to the black hole by observers hovering outside the horizon.

\item Infalling description---This is another consistent low-energy quantum mechanical description of the black hole that includes the exterior but also includes degrees of freedom in the interior of the black hole (though not all the way down to the singularity). This description does not include the stretched horizon. This is the description seen by observers falling into the black hole.
\end{itemize}
Here, the hovering and infalling observers are merely ostensive devices to generate these two descriptions; the descriptions can be understood without appeal to observers. 

With these descriptions in place, we can now state the descriptive principle of complementarity.
\begin{quote}
Descriptive complementarity: 
The exterior and infalling descriptions are descriptively consistent (as opposed to just operationally consistent) with each other and with quantum mechanics.
\end{quote}
The principle states that both descriptions can simultaneously be accurate representations of the way the world is while being consistent with quantum mechanics. While operational complementarity only required that there be no possible experimental way for the inconsistency to become salient to a single observer, descriptive complementarity takes the descriptions as representations of the world and asks if both could simultaneously be true: be taken as different ways of talking about the same world.

Something like descriptive complementarity is also endorsed by several physicists---indeed by some of the very same people who endorse operational complementarity. This shouldn't be surprising since descriptive complementarity will entail operational complementarity, but not vice versa; after all, if one has a consistent physical theory, then it won't generate any conflicting experimental implications. Some quotes: 
\begin{itemize}
\item \textcites[p. 967]{PhysRevD.49.966}: `...black
hole complementarity does not mean a departure from
the dictum that the laws of nature appear the same in
different frames of reference. Rather, the assertion is that
the \emph{description} of the same physical reality may differ
quite significantly between reference frames...' (original italics).

\item \textcites[p. 2]{lowe2016holographic}: `According to the principle of black hole complementarity... physics outside the stretched horizon of a black hole is well described by a local effective field theory but the local description does not extend inside the stretched horizon... the interior spacetime experienced by a typical observer entering the black hole in free fall is postulated to emerge from the stretched horizon degrees of freedom in a holographic fashion.' (Note: `typical observer' in this quote is eliminable; it's merely ostensive.)

\item \textcites[p. 231]{wallace_2020}: `black hole complementarity... The name invokes the noncommutativity
of quantum-mechanical observables: just as the same physical process can be described with respect to a basis of definite position or definite momentum states and may look very different in the two descriptions, so the horizon-crossing process can be described with respect to a basis appropriate to exterior physics or one appropriate to the infalling observer’s situation.'

\end{itemize}

As can be seen from the quotes above, descriptive complementarity can be thought of as a kind of holography.\footnote{See, for example,~\parencite[pp. 59-67]{raju2021lessons} and references therein.} On the holographic way of thinking, the two descriptions are just two different descriptions of the same underlying physics---essentially the interior is holographically encoded in the stretched horizon.

One point is worth emphasizing here. As suggested by the Wallace quote above,\footnote{See also, for example,~\parencites{vanDongen2004-VANOBH} and references therein.} a standard way to articulate descriptive complementarity is to require that the field operators in the exterior not commute with the operators in the interior, which means that the interior and the exterior won't have separate Hilbert spaces. This approach has the feature that, within the exterior description, information that fell into the black hole can stay inside the black hole but also be outside in the radiation---after all, the interior and the exterior are the same Hilbert space. However, this way of thinking is only be available in the exterior description and not in the infalling description. This is because the infalling description  allows smooth horizon crossings and so describes goings-on on foliations smooth across the horizon. Consequently, on the infalling description, equal-time slices across the horizon are well-defined, which means that regions inside the horizon are spacelike separated from the regions outside. So, at least on the infalling description, it cannot be that observables on the interior are non-commuting with observables on the outside, on pain of inconsistency with semiclassical physics. (This point will be essential in sections~\ref{sec:clone} and~\ref{sec:monogamy}.) 

Note that descriptive complementarity is not saying, in the manner of Bohrian complementarity about quantum mechanics, that we are only allowed to use one of the descriptions depending on one's context, that is, depending on whether one is hovering or one is infalling. This is because the restriction on the use of a description from a given context is still an operational restriction: it relies on what kind of observer one is. Further, this kind of Bohrian move does not take the descriptions representationally seriously, much like in the case of quantum mechanics.\footnote{Compare~\textcites[chapter 6]{bokulich2003horizons} on Bohrian complementarity's relation to black hole complementarity.}

\subsection{Complementarity as consistency}

Both my complementarity principles are consistency claims. These are what I am referring to as `complementarity'. Some take `black hole complementarity' to include the claim that there are the two descriptions mentioned above, or the claim that there is a stretched horizon. In this, they follow the seminal \parencites{PhysRevD.48.3743}. There, the existence of an exterior description with the stretched horizon as the boundary is the import of their three `postulates of black hole complementarity'.  

I reserve the term `black hole complementarity' for the consistency conditions that I have delineated because one does not need to appeal to an extra principle to argue for the existence of these descriptions or the stretched horizon. That much follows from low-energy field theory and general relativity. The claim that the two descriptions are consistent---operationally or descriptively---does amount to a further claim worth elevating to a principle and exploring its consequences, because what is precisely at stake here is whether quantum mechanics in the vicinity of a black hole is consistent.

In the literature, the claim about the consistency of the descriptions is often left implicit; moreover, it is not always clear whether consistency is being judged on operational or descriptive grounds. Thus, by reserving `complementarity' for the consistency claim, and by distinguishing between operational and descriptive consistency, I hope to have at least added some clarity to the discussion.

To the best of my knowledge, there has not yet been an explicit delineation between operational and descriptive principles of complementarity. The handful of philosophy-of-physics treatments of this topic (\cite[pp. 211-16]{Belot1999-BELTHI-2}, ~\cite{vanDongen2004-VANOBH}, and~\cite{Bokulich2005-BOKDBH}) have been sensitive to the operationalism implicit in black hole complementarity. However, they view operational complementarity as being useful only insofar as it is a starting point for efforts to de-operationalize it. Even if one holds this view, it is worth appreciating the breadth of the principle's applicability and the ways in which it has been employed in the recent literature.

As we shall see with the examples to follow, distinguishing the two principles will make clear the limits of descriptive complementarity, and the power of operational complementarity. I will show that the descriptive principle fails when it encounters some recent examples, while the operational principle succeeds. But first we consider an example where both principles succeed.

\section{What happens to the horizon crosser?}\label{sec:horizoncross}

Alice is an infalling observer who crosses the horizon and Bob is an observer hovering outside the black hole. Let us look at the experience of Alice as she crosses the horizon. From her perspective, the experience is smooth. She cannot distinguish her situation from traversing empty space so long as she considers physics at scales small compared to the local curvature. However, from Bob's perspective, Alice will be thermalized when she encounters the stretched horizon.\footnote{This case is from \parencites{PhysRevD.48.3743}. See also \parencites[pp. 21-2]{wallace_2020}.} This seeming disagreement between their perspectives seems intolerable, and seems to be a direct empirical ramification of the information paradox.

However, counterintuitively, there is nothing  mathematically inconsistent here. There are just these two valid descriptions and we can map one description to the other. At a semiclassical level, one can see this map as a standard coordinate change from Schwarzschild to infalling coordinates.  After all, classically, in Schwarzschild coordinates, Alice will appear to be getting closer and closer to horizon forever, whereas in infalling coordinates (such as Gullstrand-Painlev\'e), she will smoothly cross the horizon. A fully quantum mechanical treatment of mapping  the interior to the stretched horizon is much more delicate given that the stretched horizon consists of Planck-scale degrees of freedom, but steps towards such a mapping have been taken, \emph{inter alia}, by \textcites{PhysRevLett.114.201301} and \textcites{lowe2016holographic}. You don't need to buy any particular proposal for such a mapping; my aim is to simply point out that there is no obvious mathematical barrier here.

And so the infalling description is interpretable as a redescription of the exterior description, and consequently the two descriptions are descriptively consistent. Thus, descriptive complementarity holds in this case.

Let me emphasize that the kind of consistency that descriptive complementarity is talking about is mathematical consistency within descriptions and with quantum mechanics. One might doubt whether mathematical consistency is sufficient for physical equivalence---after all, it seems as if there must be a fact of the matter about whether Alice lives or dies. And, perhaps, if physical equivalence fails, then that is enough to conclude descriptive complementarity is unsuccessful. But the point here is that mathematical consistency leaves open the possibility of physical consistency. For instance, even though Alice looks to Bob as if she's been thermalized, perhaps the functional relations between microphysical degrees of freedom required for Alice to exist still obtain. However, if even mathematical consistency fails (as I will argue that it does in the following sections) then the question of physical consistency is moot. 

What about operational complementarity? Is there an observable problem here? Perhaps Alice can provide evidence to Bob that she was not thermalized at the stretched horizon. If she sends a signal to Bob saying she's fine after crossing the stretched horizon, then Bob will know something is wrong with his model of the black hole. For he would then both see Alice thermalized at the stretched horizon and have confirmation that Alice safely crossed the stretched horizon.\footnote{Being operationalist about complementarity---or quantum mechanics---allows any observer to still treat other observers as physical systems: `Nothing in principle prevents us from quantizing a colleague'~\parencites[p. 70]{fuchs2000quantum}.   So for an operationalist, Bob can treat Alice as a quantum system.}

The trouble is that since the stretched horizon is only one Planck length above the true horizon, Alice must send her message using field modes of Planck-length frequencies (if it even makes sense to talk about `field modes' and `frequencies' at that scale). Otherwise, it'll be too late: she would have crossed the horizon and her message can't get out. Conversely, if Bob wants to probe what is happening to Alice as she crosses the stretched horizon, he needs to send in modes of Planck-scale frequencies to be able to resolve what is happening. Neither can he hover close to the horizon to examine near-horizon physics\footnote{He will definitely have to hover. He cannot, for example, just enter an orbit that gets him close to the horizon because there are no orbits around a Schwarzschild black hole (bound or unbound) below 1.5 Schwarzschild radii. See, for example,~\parencites[Box 25.6]{misner1973gravitation} for details.}---to do so would require sustaining Planck-scale accelerations.\footnote{Classically, the proper acceleration of a hoverer at the horizon diverges.} 

For this case, then, both descriptive complementarity and operational complementarity succeed because neither are we able to locate a descriptive contradiction nor an operational contradiction. This changes in the more involved scenarios that we now turn to.

\section{Quantum cloning?}\label{sec:clone}

\begin{figure}[!t]
\centering
\includegraphics[width=0.33\columnwidth]{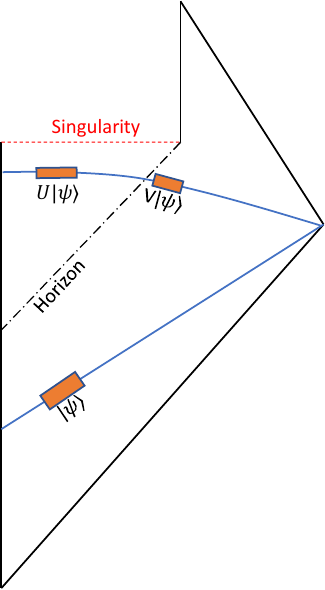}
\caption{Penrose diagram depicting that the quantum information in the infalling matter $\ket{\psi}$ is cloned. That is, unitary transformations of it, $U\ket{\psi}$ and $V\ket{\psi}$, are generated in the interior and in the Hawking radiation in the exterior.}
\label{fig:cloner}
\end{figure}

Consider some quantum information encoded in infalling matter. Now, from the perspective of infalling observers, this information uneventfully crosses the horizon and continues on towards the singularity. However, this information will become accessible to exterior observers in the radiation after the Page time (see section~\ref{sec:bhip}). As the radiation state becomes purer and purer (following the Page curve), the quantum states of degrees of freedom that fell in earlier will become recoverable from the radiation. This suggests that the quantum information that was present in the infalling matter has been cloned: a copy of what's in the interior is also in the radiation.\footnote{This was first noted in~\textcites{PhysRevD.49.966} and significantly sharpened by~\textcites{Hayden_2007}.} This looks like a violation of the no-cloning theorem of quantum mechanics, which states that no unitary process (indeed, no linear process) can copy an arbitrary quantum state.\footnote{See, for example,~\cite[pp. 24-5]{nielsen2002quantum} for review of the no-cloning theorem).} In our case, the no-cloning theorem says that there cannot be a unitary transformation connecting quantum states on a slice with the infalling matter before it fell in and a slice that contains (unitary transformations of) both the quantum state of the fallen-in matter and the same quantum state in the radiation (see Figure~\ref{fig:cloner}). However, it seems this must indeed be the case if black hole evaporation is unitary.

In this scenario, descriptive complementarity fails. Taking both the exterior and infalling descriptions representationally seriously, we have an inconsistency with quantum mechanics. To see this, note that the exterior description will say that Alice's clone exists in the exterior, in the radiation. Moreover, in the exterior region, the exterior and infalling descriptions agree. But on the infalling description, Alice exists in the interior as well. On the infalling description, one can construct spacelike slices that are smooth across horizon, because on this description there is no stretched horizon and the horizon is locally indistinguishable from empty space. Now, because the infalling description contains Alice in the interior and Alice in the exterior (by virtue of its agreement with the exterior description), we can define smooth slices which contain clones. Consequently, we have, upon radiation of the relevant degrees of freedom, a quantum cloning process. This patently violates quantum mechanics, and hence we must conclude that descriptive complementarity fails. 

Some attempt to avoid the cloning problem while maintaining a descriptive picture by providing a reason why one is not allowed to consider slices that are smooth across the horizon. For instance, \textcites{PhysRevD.48.3743} say that because there are no `superobservers' who can conduct experiments both in the interior and the exterior, we cannot compare descriptions pertaining to those regions. There are two issues with this line of argument. First, this is an explicitly operational restriction because it appeals to capabilities of observers. As such it won't help rescue descriptive complementarity. Second, it is not quite right that there no observers capable of observing the interior and exterior. It's true that interior observers cannot escape to the exterior, but nothing stops observers initially in the exterior from entering the interior. Indeed, below we consider precisely such cases. Now, it is true that there are no observers who can simultaneously conduct experiments in the interior and the exterior. But that can't be reason enough to discount slices connecting the two regions. After all, for any two spacelike separated regions, no observer can truly simultaneously conduct experiments in both regions. Either an observer conducts an experiment in one region and then travels to the other or two observers conduct experiments in both regions and subsequently meet up to compare notes. And both options are still available even with a horizon in between: an experimenter can start in the exterior and then go to the interior, or meet up with someone already in the interior. 

So can an exterior observer enter the interior and observe the clone? If so, operational complementarity would also fail. To better see how this might work, let's return to infalling Alice and hovering Bob. Say Alice carries with her a quantum bit (or a qubit) as she falls into the black hole. Bob waits, patiently collecting Hawking radiation, until, past the Page time, the information that Alice carried in reappears in the radiation. He then jumps into the black hole. Alice then  sends her qubit to Bob (whom she knows will jump in) via a photon. If Bob can intercept Alice's photon and compare it with the information he recovered from the radiation, then it seems he can directly see a clone, and thus detect a violation of the no-cloning theorem.\footnote{This can be done, for instance, if they pre-decide that Alice will carry the quantum state $\ket{\psi}$; then all Bob has to do is measure with the projectors $\{\ket{\psi}\bra{\psi}, I - \ket{\psi}\bra{\psi}\}$.}

\begin{figure}[!t]
\centering
\includegraphics[width=0.4\columnwidth]{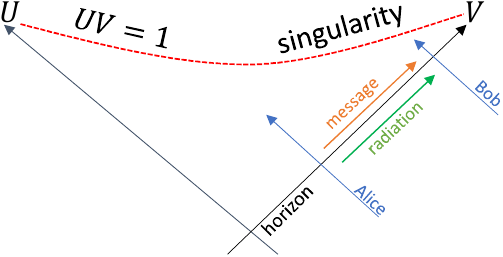}
\caption{$U$ and $V$ are Kruskal coordinates. Alice has to send her information to Bob before Bob hits the singularity. (Adapted from \parencites[Figure 2]{Hayden_2007}.)}
\label{fig:kruskal}
\end{figure}

However, if operational complementarity is right, then there will be an operational barrier for this experiment. And indeed, there is. For this proposal to work, Bob must receive the message from Alice before he crashes into the singularity.  The longer Bob waits to jump in, the shorter Alice has after she crosses the horizon to send her message to Bob if he is to receive it before he hits the singularity (see Figure~\ref{fig:kruskal}). This is the source of the operational barrier. More precisely, \textcites{PhysRevD.49.966} argued that Alice only has an extremely short time to send her message: it must be sent within a time-scale exponentially small in the square of the black hole mass so as to reach Bob before he encounters the singularity. So she must encode her information in a signal of frequency exponentially large in the square of the black hole mass. Thus, unless Alice can access Planck-scale frequencies, she cannot send a message to Bob that will allow him to verify a violation of no-cloning. Thus, operational complementarity is vindicated. 

But wait. Perhaps operational complementarity only seems to be vindicated because we weren't clever enough. For there is a significant weakness in \textcites{PhysRevD.49.966}'s argument: they assumed Bob needs to wait for as long as the Page time before Alice's information comes back out in the radiation. They needed to assume this because, as discussed in section~\ref{subsec:page}, the radiation before the Page time will not be sufficient to allow Bob to reconstruct Alice's qubit. And the Page time is very long: it is of the order of the black hole evaporation time. However, \textcites{Hayden_2007} showed that if information falls in after the Page time, then black holes re-emit that information very quickly. Indeed, they showed that such information comes back out on a time-scale of the order of $M \log M$ (in Planck units). For solar mass black holes, this is about $10^{-4}$ seconds! This is extremely short compared to the evaporation time or the Page time of such black holes, which is of the order of $M^3$, which, for solar mass black holes, is about $10^{63}$ years. Thus, Hayden and Preskill call old black holes---that is, black holes past their Page time---`information mirrors', because after the Page time, any information you throw in comes right back at you.

This restores hope that there might indeed be a way for Bob to directly observe cloning. If Alice jumps into a black hole after the Page time, then, as argued above, Bob receives the information in Hawking radiation soon after she jumps in; following which, he jumps in as well. But strikingly, \textcites{Hayden_2007} show that even in this most forgiving scenario, Alice still does not have enough time to send the signal after her horizon-crossing and have it reach Bob before he hits the singularity. They showed that if Alice sends the signal after a time that is longer than $\mathcal{O}(M \log  M)$, then it would be too late for Bob to receive the signal. But that is precisely the time-scale for which Bob has to wait before he has to jump in if he wants to recover Alice's information from the radiation! More carefully, the time difference between when Alice has to send the message and when Bob has to jump in is of the order of the Planck time, meaning that Alice has to encode her qubit in Planck-scale modes, preventing Bob from seeing the clone. Thus we see that operational complementarity really is vindicated.

The case quantum cloning provides a clear example wherein operational complementarity succeeds but descriptive complementarity fails. Let me emphasize, however, that I don't mean to suggest that those who worked out descriptive complementarity proposals were making some sort of elementary error. Obviously not. Rather, what seems to be happening there---as I said in my discussion of appeals to `superobservers'---is that such proposals seem to contain some implicit appeal to operational constraints. So part of my point is that descriptive strategies---taken seriously as descriptive-representational strategies---fail on those terms, but might be defensible once operational notions are imported. So we should lean more towards accepting, if only provisionally, operational complementarity, even if we are uncomfortable with operational principles in our physics.

Before moving on, it's worth considering why the problem of cloning discussed in this section could not have been raised in the previous section, with the scenario of the horizon crosser. In that case too, doesn't the information in the degrees of freedom constituting Alice get encoded in the stretched horizon after she crosses the horizon, resulting in clones located in the interior and on the stretched horizon? But there, the descriptive complementarist can respond that the description of Alice as smeared on the horizon is not part of the exterior description for it is not a low-energy description. More precisely, we don’t really know if we can describe the way in which Alice is encoded on the stretched horizon as a quantum field state localized on a certain spacelike slice. At the Planck scale, we may not even be able to distinguish between spacetime and field-states on spacetime.\footnote{This is the case in some promising ideas to derive general relativity as an emergent thermodynamical theory from quantum degrees of freedom. See, for example, \textcites{van2010building,PhysRevD.86.065007,PhysRevLett.116.201101,PhysRevD.95.024031} On these proposals, the structure of spacetime arises from the structure of interactions between Planck-scale quantum degrees of freedom, and so it's not clear that they spacetime and field states are well-defined notions at the Planck-scale. \label{fn:sptbreak}} However, once Alice's information appears in the radiation, then a low-energy description of that information becomes available, allowing us to identify a problem for descriptive complementarity. For similar reasons, the operational complementarist will be untroubled by the appearance of cloning before radiation: the clone on the stretched horizon isn't empirically accessible to observers who can't probe Planck-scale degrees of freedom.

\section{Violating entanglement monogamy?}\label{sec:monogamy}

Consider another potential experiment intended to set up a violation of quantum mechanics observable by a single observer: the famous AMPS paradox (named after its discoverers Almhieri, Marolf, Polchinski, and Sully \parencites{almheiri2013black}; sometimes called `the firewall paradox'). Suppose, again, we have unitary evaporation. This entails that the early Hawking radiation is going to be near-maximally entangled with the late (that is, post-Page-time) radiation. This must be the case because the late radiation starts purifying the total state of the radiation, as seen in the Page curve. Now say Bob does the following. He collects all the early radiation until after the Page-time. Then goes close to the horizon and collects the late radiation that ought to be near-maximally entangled with all the early radiation, as predicted by the Page curve. Given the large amount of entanglement between the early and the late radiation, he should be able to distill, from all the radiation that he has collected, a quantum state close to a pure state. But we also expect that the modes near the horizon---that is, the late radiation we just collected---will be highly entangled with modes just behind the horizon, since Hawking radiation arises from tracing over the highly entangled vacuum of a relativistic quantum field theory. However, the principle of monogamy of entanglement says that the same quantum system cannot be highly entangled with two different systems.\footnote{See, for example, \parencites[p. 917]{RevModPhys.81.865}.} So it seems Bob could observe a violation of the monogamy of entanglement by distilling a large amount of the entanglement between the early and late radiation into a pure state, and then crossing the horizon and checking if the resultant state is still entangled with modes behind the horizon. Thus, it seems a single observer can observe a violation of quantum mechanics.

This thought experiment results in a failure of descriptive complementarity. We have large amounts of entanglement between the late radiation and the early radiation, while also having large amounts of entanglement between the late radiation and interior modes. All these three quantum systems can be located, according to the infalling description, on a single spacelike slice that smoothly traverses the horizon (see section~\ref{sec:clone}). This violates monogamy. The exterior and infalling descriptions taken descriptively seriously, contradict quantum mechanics.

Operational complementarity continues to succeed. \textcites{harlow2013quantum} argued that if Bob attempts to perform the AMPS experiment, he will fail because the task of distilling the entanglement between the early and late radiation will take much longer than the evaporation time of the black hole, thus destroying any modes behind the horizon that would allow us to observe a violation of monogamy. The argument for this is based on computational complexity theory. A strengthened version of Harlow and Hayden's argument was given by Aaronson,\footnote{Aaronson hasn't published this argument; see~\parencite[pp. 48-9]{RevModPhys.88.015002} for a version in print.} who showed that if the task of distilling the Hawking radiation could be performed efficiently---that is, in a time that is polynomial in the entropy of black hole---then a complexity-theoretic conjecture that is widely believed to be true, and widely employed in the security proofs for cryptographic protocols, would be false. Therefore, Bob cannot see a direct violation of entanglement monogamy, for the black hole would finish evaporating before he is ready to jump in and compare his distilled state with the modes behind the horizon.

Could Bob get lucky and obtain the correct distillation without running an exponentially long computation? Sure. However, he wouldn't be able to verify that he lucked out in this way without also destroying the distillation he obtained: quantum measurements are typically destructive. And if he destroys the distillation then he can't observe the violation of monogamy. And he couldn't have made a copy of the distillation since that's blocked by the no-cloning theorem. The only real way he can know that he got the right distillation is by relying on a quantum computer to work properly. Perhaps there's some clever way he can by sheer luck hit upon the correct state and non-destructively verify that he did so, but the details of such an idea would have to be developed and defended, and that would then take us right into the kind of dialectic that operational complementarity engenders.

Maybe Bob could collect and manage all the physical processes that occur during the final phase of evaporation and use that to check against his computation that monogamy really was violated? But that would require he access Planck-scale physics, for in the final stages of evaporation, the horizon intersects with the singularity.

(It's worth remarking here that the fact that Aaronson's argument relies on a conjecture is no mark against it. Computational complexity\footnote{See, for example, \parencites{arora2009computational} for an introduction.} is a very successful field, with important practical applications, but much of it relies on conjectures---the most famous of which is $P\neq NP$---that while unproven, are widely believed to be true, and hence frequently relied upon in proofs. See \parencites{Aaronson2016} for a review of progress on and barriers to proving the $P\neq NP$ conjecture, and a defence of why one ought to think it true. See \parencites{Williams2019} for a  discussion of many conjectures in computational complexity and their plausibility.)

So we see that operational complementarity is secure. We are unable to identify an operational contradiction in this scenario. Attempting to directly observe the violation of monogamy is foiled by computational complexity. The fact that the barrier is computational highlights the value of a truly operational principle here.

Harlow and Hayden's result has been influential in the field. But there have been attempts to circumvent the computational barrier they identify, for example, by~\textcites{oppenheim2014firewalls}. However, further barriers to such attempts have been identified~\parencites{RevModPhys.88.015002, aaronson2016complexity,kim2020ghost}. And even if these barriers fail, other barriers remain~\parencites{bao2016rescuing,yoshida2019firewalls}.

Given the various proposals and counter-proposals and counter-counter-proposals that look for, or attempt to refute, operational barriers to verifying violations of quantum mechanics, it is clear that operational complementarity is crucial in guiding this discussion. Thus, given the literature so far, operational complementarity remains a plausible and powerful principle. 

\section{Lessons}\label{sec:concl}

An operational formulation of black hole complementarity has been essential in the recent literature surrounding the black hole information paradox. This literature shows that attempts to extract, from the paradox, observable violations of quantum mechanics in above-Planck-scale physics fail. Very promising proposals to generate observable violations of quantum mechanics are thwarted for subtle reasons. Meanwhile, a descriptive version of complementarity is unsuccessful:  the exterior and infalling descriptions taken representationally seriously conflict with quantum mechanics. 

So where do we go from here? For scientific realists, the failure of descriptive complementarity is perhaps none too surprising. After all, the black hole information paradox identified an inconsistency in the application of quantum mechanics to a black hole. Given an inconsistency, it is no surprise that the inconsistency reappears in different guises in different thought experiments. You can run but you can't hide.

Nevertheless, realists ought to be surprised by the success of operational complementarity. The fact that the inconsistency cannot be ramified up to an experimental problem likely signals something about the deeper descriptive theory that would resolve the information paradox. This is analogous to how, in the case of special relativity, the inability of observers to agree on which events are simultaneous signals the geometry of Minkowski spacetime. Or how, in quantum mechanics, the inability of observers to simultaneously measure precise values of position and momentum signals the nature of the wavefunction. Thus, even realists must take seriously the success of operational complementarity for it provides both clues towards, and new explananda for, future physics. 

Thus, those of realist inclination should view operational complementarity as a way station en route to a fully descriptive theory, one that accounts for the information paradox in a way that makes no mention of observers. If one adopts this attitude, the Bell-style objection to operationalism discussed in section~\ref{subsec:opdescpcomp})---that observers are physical systems like any other and hence should not appear as primitives in an adequate physical theory---is taken seriously, but the job of providing a satisfactory response to it is deferred to future physics. Structurally, this is similar to the attitude that those who seek realistic interpretations of quantum mechanics take towards the success of the quantum recipe (that is, unitary evolution + the Born rule).

On the other hand, for those with no objection to operationalism in physics, the success of operational complementarity indicates that the black hole information paradox has been resolved. For what is a paradox? It's a compelling argument to an absurd conclusion. If one is operationalist, then the absurdity of the conclusion has to be cashed out in operational terms. The success of operational complementarity suggests that there is no operational absurdity arising from the black hole information paradox---no experiment can be done to bring out the contradiction. Consequently, for operationalists, as long as future work doesn't invalidate operational black hole complementarity, there's no paradox left.









\section*{Acknowledgements}
I am indebted to David Wallace for countless enlightening discussions and comprehensive feedback on drafts. I am grateful to Ignacio J. Araya for innumerable conversations on the physics of black holes. Thanks to Sam Fletcher, Bixin Guo, John Norton, and especially, Nick Huggett for helpful comments on drafts. And thanks to Gal Ben-Porath, Daniele Oriti, Renato Renner, and Aron Wall for valuable comments and discussions. I also thank two anonymous referees who made several constructive suggestions which significantly improved this paper. Thanks also to audiences at the Philosophy of High-Energy Physics conference at the University of Pittsburgh, at the Philosophy of Science Association meeting in Pittsburgh, at the XXV Urbino Summer School, at the Foundations of Gravity Conference at LMU Munich, and at the LMP Graduate Conference at Western University.



%

\printbibliography[title ={References}]

\end{document}